# 异常信息敏感的框架 API 生命周期模型构造*

燕季薇 [1),2)]  黄进豪 [3)]  杨恒钦 [2),4)]  严俊 [1),2)]*

[1)](中国科学院软件研究所 软件工程技术研究开发中心 北京 100190)

[2)](中国科学院大学 北京 100049)

[3)](北京工业大学 北京 100124)

[4)](国科大杭州高等研究院 杭州 310024)

**摘　要**　大型软件系统的实现通常依赖于底层框架或第三方库。这些框架/库代码数量繁多、实现复杂，但它们的演化升级往往独立于其调用者，为上层软件的质量保障带来挑战。例如，框架/库代码升级时 API 的新增和删除行为可能引发上层软件的兼容性问题。准确分析并提取框架/库代码 API 生命周期模型可以有效辅助对这类代码演化情况的理解，支撑上层的分析与测试。现有工作中的 API 生命周期模型主要关注 API 的存在性变动，而未考虑特定代码语义变更对开发者的影响。在某些情况下，开发者没有删除或废弃特定的 API，而是更改其代码语义，例如增加、删除或修改用于外部数据校验的异常抛出相关代码。如果新版本框架/库中的异常抛出行为发生变更而调用者不知情，可能给上层软件系统带来隐患。因此，除了 API 的存在性，开发者还应特别关注异常相关代码的变更情况，即为框架/库代码构建异常信息敏感的 API 生命周期模型。

　　为实现异常信息敏感的 API 生命周期模型构造，本文采用面向 Java 字节码的静态分析方法，首先提取框架 API 中的异常抛出行为，生成异常摘要信息，然后通过多轮流式匹配策略获取异常信息的变更情况，构造异常信息敏感的 API 生命周期模型。该方法：1）通过控制依赖语句切片提取异常抛出语句的关键触发条件，采用参数推断策略将局部变量的约束条件转换为仅与外部输入参数相关的异常前断言，并基于自底向上的摘要传递实现跨过程异常摘要提取；2）通过关键信息精准匹配和自适应模糊匹配策略，分析异常摘要信息的新增、删除和修改情况，最终得到异常敏感的 API 生命周期模型，共包含七种 API 变更形式。基于该方法，实现了基于 Java 字节码分析的 API 生命周期提取工具 JavaExP。与最新的 Java 异常分析工具相比，在异常摘要信息提取方面，JavaExP 在大幅提高分析效率（+7x）的同时实现了更高的 F1 分数(+60%)。通过人工确认 Apache common-io 项目在 19 个版本上的演化报告，发现异常级别的演化分析准确率达到 98%。对六个真实框架/库项目在 60 个版本上的 API 生命周期分析表明，与异常不敏感的 API 生命周期相比，采用异常敏感的分析时，API 发生变动的比例提高了 18%。在 75,433 个被分析的 API 中，约有 20% API 的异常抛出行为至少发生过一次改变，这些 API 共涉及超过七千多处独立的异常变更。在多个项目上的分析结果表明，异常敏感的模型构造能够更加精准地描述 API 的演化过程。

**关键词**　静态分析；代码演化；Java 异常摘要；API 生命周期

**中图法分类号**　TP311　**DOI 号**　xxx

# Exception-aware Lifecycle Model Construction for Framework APIs

YAN Ji-Wei[1),2)]   HUANG Jin-Hao[3)]   Yang Heng-Qin[2),4)]   YAN Jun[1),2)]*

[1)](Technology Center of Software Engineering, Institute of Software, Chinese Academy of Sciences Beijing, 100190)

[2)](University of Chinese Academy of Sciences, Beijing, 100049)

[3)](Beijing University of Technology, Beijing, 100124)

[4)](Hangzhou Institute for Advanced Study, Hangzhou, 310024)

**Abstract**

The implementation of complex software systems usually depends on low-level frameworks or third-party libraries. However, the evolution of these frameworks or libraries is independent of the upper-level applications, which brings challenges in upper-level code quality assurance. For example, during code evolution, the APIs adding and removing behaviors may cause unexpected compatibility problems. Precisely analyzing and constructing the framework/library's API lifecycle model is of great importance, which could help in understanding changes in APIs as well as supporting the analysis and testing of upper-level code. Nowadays, existing works propose the API existence-changing model for defect detection, while not considering the influence of semantic changes in APIs. In some cases, developers will not remove or deprecate APIs but modify their semantics by adding, removing, or modifying their exception-thrown code, which is used to verify the users' inputs. It may bring potential defects to upper-level code if the exception-related behaviors in newer versions are changed silently. Therefore, besides the API existence model, it is also necessary for developers to be concerned with the exception-related code evolution in APIs, which requires the construction of exception-aware API lifecycle models for framework/library projects.

To achieve automatic exception-aware API lifecycle model construction, this paper adopts static analysis technique to extract exception summary information in the framework API code and adopts a multi-step matching strategy to obtain the changing process of exceptions. Then, it generates exception-aware API lifecycle models for the given framework/library project. Our approach: 1) adopts control-dependency slicing analysis to extract the conditions of the exception-thrown statements; uses a parameter tracing strategy to transform exception-throwing conditions into external-variable-related preconditions; and performs inter-procedure precondition construction by a bottom-up summary-based analysis. 2) proposes the exact-matching and adaptive-matching strategies to analyze the addition, deletion, and modification changes based on the summarized exception summaries; generates exception-aware API lifecycle model which covers seven API changing types. With this approach, the API lifecycle extraction tool, JavaExP, is implemented, which is based on Java bytecode analysis. Compared to the state-of-the-art tool, JavaExP achieves both higher F1-score (+60%) and efficiency (+7x). By manually confirming the exception changing reports on 19 versions of Apache common-io project, we found that the precision of exception matching and changing results is 98%. The evaluation of 60 versions of six projects shows that, compared to the exception-unaware API lifecycle modeling, JavaExp can identify 18% times more API changes. Among the 75,433 APIs under analysis, 20% of APIs have changed their exception-throwing behavior at least once after API introduction. These APIs involve a total of more than 7k independent exception changes, which shows that the exception-aware lifecycle modeling can describe the evolution process of APIs more accurately.

# 1. 引言

具有复杂功能的大型软件系统往往由多个模块组成，其实现依赖于底层编程框架和种类繁多的第三方库。这些框架/库代码通过持续的版本更新修改代码缺陷或完善代码功能，其演化过程独立于调用它们的上层软件系统。在上层应用的开发过程中，软件供应链安全分析中的依赖安全检测工具会帮助开发者识别项目依赖中的漏洞，并提醒应用开发者尽快更新版本以保障代码质量安全。例如，当 GitHub 检测到项目代码中使用易受攻击的依赖项或恶意软件时，会向开发者发送 Dependabot 警报[1]。如果上层应用开发者在不熟悉框架/库代码 API 演化过程的情况下变更版本，可能会引入其他问题，如使用了过时/被移除的 API 或未及时捕获处理新抛出的异常，进而导致程序错误或引发兼容性问题等。

框架/库代码中 API 的变更导致用户在迁移上层软件系统时花费较多精力，间接增加了使用特定框架/库的开发难度[2][3][4]。为保障这些依赖于底层框架/库函数的软件系统质量，一种解决方案是预先分析不同框架/库代码版本下的 API 调用规约，并检测调用代码的正确性。针对这一问题，现有工作提出了 API 级别的生命周期模型[18]，并通过分析框架更新时提供的 API 变更文本文件[5]或通过轻量级框架代码分析扫描其 API 列表[6][7]等方法来构建 API 级别的生命周期模型。虽然这些工作考虑了 API 的存在性变更，但忽略了 API 中关键代码语义变更的影响。在某些情况下，开发者没有删除或废弃特定的 API，而是更改其代码语义。对于框架/库的使用人员，除了 API 的存在性变更外，语义信息变更也是 API 调用时以及 API 调用合规性检测中需要考虑的一项关键信息。

基于 API 完整代码的差异分析可以准确反映代码的变更情况，但完整的代码变更结果数据量庞大且复杂，对于上层应用的分析和测试，并非所有变更都会对用户的使用产生影响。我们发现，在 API 演化过程中，同一 API 的基本功能往往保持一致，即代码升级不应影响现有 API 的基本功能实现，这类变化应是对用户透明的，但 API 对外部输入数据的校验过程和校验结果的反馈方式是可变的，它们会影响 API 的上层调用。在 Java 代码中，当 API 接收到非预期的外部输入时，通常会抛出异常来应对这一非预期行为，而上层用户应该及时捕获并处理这些异常行为[41]。Mostafa 等人对 Java 库代码兼容性错误的统计[55]表明，由异常导致的兼容性问题占比超过 1/3（105/296）。葛等人在文献[58]中指出，框架/库代码中存在的错误或漏洞可能会被攻击者利用，从而损害软件供应链安全，这些错误或漏洞往往与框架/库代码中存在的异常有关。由此可见，API 异常抛出行为的变化会对用户调用方式产生重要影响，在软件快速演化背景下，其变更行为对于软件的健壮性与安全性息息相关。

```
1. public int getCount() {
2. -    return (int) getByteCount();
3. +    long result = getByteCount();
4. +    if (result > Integer.MAX_VALUE) {
5. +        throw new ArithmeticException("The byte count " + result + " is too large to be converted to an int"); }
6. +    return (int) result;    //修改返回空值为抛出异常
7. }
```
**（a）新增异常实例**

```
1. public static void moveFile(File srcFile, File destFile) throws IOException {
2.     if (destFile.exists())
3. -        throw new IOException("Destination '" + destFile + "' already exists");
8. +        throw new FileExistsException("Destination '" + destFile + "' already exists"); //修改异常类型
4. }
```
**（b）修改抛出异常类型**

```
1. public void forceDelete(File file) throws IOException {
2. +    boolean filePresent = file.exists();
3. +    if (!file.delete()) { //增加文件删除判断条件
4. +        if (!filePresent) {
5. -        if (!file.exists()) {
6.             throw new FileNotFoundException("File does not exist: " + file); }
7. +    }
8. }
```
**（c）修改异常抛出条件**

**图 1 不同版本 API 中异常相关代码变更示例**

图 1 给出了真实项目中的异常相关变更代码片段示例，其变更形式多种多样，包括新增或删除异常实例[12][13]、修改异常实例的类型、描述或抛出条件[14][15][16]等。在代码演化过程中，同一异常可能发生多次不同类型的变更[14][17]（参见图 3）。这些变更可能对 API 的外部使用产生影响，即改变

API 使用规约。因此，为了正确理解框架 API 的生命周期行为,需在考虑 API 增删变化之外,结合 API 中异常的变更情况，为其构造异常信息敏感的生命周期模型。该模型的构建依赖于对异常摘要信息的准确提取与匹配分析。其中的关键挑战是：1）应提取哪些关键信息表征 API 中的异常实例并尽量减少数据中影响匹配的噪音；2）当一个方法中存在多个同类型的异常实例时，如何准确地在多个版本中匹配到同一实例并识别变更内容，从而准确构建 API 及其异常集合的生命周期。

针对这些挑战，本文首先设计了一种面向演化分析的异常摘要形式，包括异常类型、描述文本、前断言三类核心信息。为了减少匹配时的数据噪音，本文在异常抛出条件分析中，通过控制依赖约束分析去除了与异常抛出无关的条件约束；通过数据流分析将所有中间局部变量约束转换为外部输入变量约束；通过跨过程异常传递分析避免函数级代码重构导致的函数内异常变更。此外，为准确识别不同版本代码中的异常实例，本文共设计了基于类型、描述、前断言、关键前断言四类信息的过滤器，对于无法完全匹配的异常实例，采用多轮流式匹配策略识别变更实例、分析变更过程，最终生成涵盖七类变更行为的异常敏感 API 生命周期报告。

基于该方法，本文实现了异常信息敏感的 Java API 分析工具 JavaExP (Java Exception-aware API analyzer)[46]。其框架图如图 2 所示，对于任意两个（或多个）版本的 Java 框架/库的 jar 包/class 文件，JavaExP 先通过异常摘要提取模块获取每个版本的异常摘要报告。其次，将这些摘要被输入到生命周期构造模块，对不同版本中 API 异常语义摘要执行自适应匹配和异常变更分析。分析后,可以获取 API 的新增、删除情况和 API 中异常的新增、删除和修改情况，从而得到目标版本区间上异常信息敏感的 API 生命周期模型。

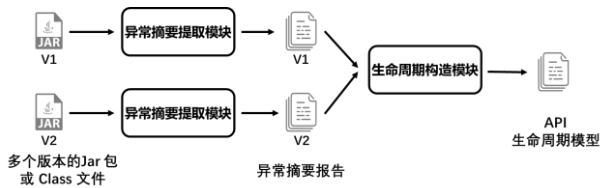

图 2 JavaExP 方法框架图

多组对比实验验证了本文方法的有效性。对于异常摘要提取模块，与最新的 Java 异常分析工具 WIT [26]相比，JavaExP 使用更短的时间 (-87%) 提取到了大量 WIT 无法分析的异常信息，并显著提高 F1 分数（相对提升 60%）。应用 JavaExP 分析了六个项目的 60 个版本，并为其生成 API 生命周期报告，演化分析的准确率达到 98%；找到了 API 中大量的异常变更行为，在 75,433 个 API 中，在异常敏感的 API 生命周期模型中，约 20%的 API 在首次引入后，异常信息发生过至少一次变动；与异常不敏感的 API 生命周期相比，异常敏感的 API 发生变动的比例提高了 18%。本文的工具和实验数据均已开源到 GitHub [46]。

本文的章节结构设计如下：第 1 章概述异常信息敏感的 API 生命周期模型构造方法；第 2 章介绍本文所需的基础知识和概念定义；第 3 章介绍面向 Java 程序的异常摘要提取方法；第 4 章介绍基于异常摘要分析 API 生命周期构造方法；第 5 章给出实验设计和结果分析，评估所提方法的有效性；第 6 章介绍本文的相关工作；最后一章为总结与展望。

## 2. 基础知识与示例应用

### 2.1. Java异常

异常是在程序执行过程中出现的问题或错误的一种表示。在 Java 语言中，异常被定义为派生自 java.lang.Throwable 类的对象，在 Java 类库、用户方法及运行时故障中都可能会抛出异常。Java 提供了很多内置的异常类 [42]，如 IOException、IllegalArgumentException 等，此外，开发人员还可以自定义异常类以便更好地适应特定需求。一部分异常会被 Java 虚拟机自动的抛出，在运行时不需要显式处理，但它们可能会导致程序的异常终止，这类异常也被称为非受检异常或运行时异常；还有一类异常需在编译时显式地处理，否则会导致编译错误，它们又叫受检异常或编译时异常[43]。在方法内部检测到不符合预期条件或无法处理的情况时，开发者可以通过 throw 语句声明主动抛出异常提供有关特定问题或错误的信息，并将控制权交给调用者或上层代码来处理（try-catch 行为）。对于框架/库 API 的调用者，API 中抛出的异常类型、抛出条件等与 API 调用过程的中数据输入规约和异常捕获方式息息相关。因此，这类变更应被及时传递给开发人员。

### 2.2. API 演化和生命周期

应用程序接口（API）是框架/库代码对外提供服务的调用接口。随着代码功能的演化升级，旧版本

API 在新版中可能被删除，新版代码中也会增加 API 以提供更丰富的功能。除了 API 的增加与删除，API 中代码的实现方式、数据校验方式等均可能发生变化。API 的生命周期指 API 在不同的框架/库版本中的存在范围，如 Li 等人提取了安卓框架代码中 API 的生命周期模型[5]，虽然 API 的变动较为频繁，但在 API 持续存在的生命周期中，其中包含的异常实例可能是持续演化的，这类变化在异常不敏感的 API 级别生命周期模型中无法体现。

**2.3. 概念定义**

针对 API、异常及生命周期等概念，我们分别给出如下定义。

**定义 1（应用程序接口方法）：** API = (id, version, class, method, S_Exp) 为一个应用程序接口方法，它包含 API 的签名、版本号、所在的类名称、方法名称以及包含的异常集合 S_Exp。

**定义 2（异常摘要）：** Summay(exp) = (API.id, type, message, condition, precondition), exp ∈ API.S_Exp 为一个异常摘要，它包含异常所在的 API、异常的类型、异常抛出时的描述文本信息、异常抛出语句的控制依赖条件以及与异常抛出相关的外部参数前断言。

**定义 3（API/异常存在性生命周期）** API 的存在性生命周期为该 API 从引入到删除的版本区间的并集。对于其中包含的异常实例 Exp∈S_Exp，异常的存在性生命周期为该异常在该 API 中从引入到删除的版本区间的并集。

**定义 4（异常敏感的 API 生命周期）** API 中所有异常对象 exp∈API.S_Exp 摘要信息 Summay(exp)的集合为 S_Summay(S_Exp)，在不同的版本中，当且仅当两个异常摘要信息集合中任何异常摘要信息均相同时，可认为异常摘要信息的取值在不同版本上保持不变。对于异常摘要信息集合的特定取值，如果其出现的最早版本为 $V_i$，最末版本为 $V_j$ (i<=j)，则其生命周期为[$V_i$, $V_j$]。对于异常敏感的 API 生命周期，它给出了摘要信息集合 S_Summay 在目标版本中所有不同取值到其生命周期的映射关系。导致 API 中异常摘要信息取值变更的操作被称为异常敏感的 API 操作，包括：API 新增、API 删除、API 修改-异常新增、API 修改-异常删除、API 修改-异常类型变更、API 修改-异常描述变更和 API 修改-异常断言变更七类。

**2.4. 示例代码**

图 3 为 Apache commons-io [14][17]项目中 API moveFile()的部分代码，给出了其异常 e 在不同版本中的变更情况，+表示新增代码，-表示删除代码。该异常为针对 API 第 2 个参数变量 destFile 的文件存在性校验。在版本号 1.4 的代码中，类型为 IOException 的异常实例 e 被引入，而在 2.0 版本中，异常实例 e 的类型被更改为 FileExistsException 类。随后，在版本 2.7 中，该方法被重构，但实际异常 e 的抛出条件未发生变化。在版本 2.9 中，该方法被再次重构，异常 e 的抛出位置发生变化，其描述文本和异常前断言也发生改变。因此，对于图 5 中的异常 e，其演化过程为一次新增，一次异常类型更改，一次异常描述文本和前断言更改，其中经历了两次方法重构。在这种情况下，上层开发者很难快速评估不同版本中 API 包含的异常是否变动以及变动的过程。为提取目标 API 的生命周期，首先需准确获取每一个异常实例的关键信息，判定不同版本中的多个异常信息是否指向同一个异常实例，如是同一实例，则记录其变更过程，在此基础上，生成完整的 API 生命周期报告。

```
1.  public static void moveFile(File srcFile, File destFile) ...{
2.    if (srcFile == null) {
3.        throw new NullPointerException("Source must not be null");
4.    if (destFile == null) {
5.        throw new NullPointerException("Destination must not be null");
6.    if (!srcFile.exists()) {
7.        throw new FileNotFoundException("Source '" + srcFile + "' does not exist");
8.    if (srcFile.isDirectory()) {
9.        throw new IOException("Source '" + srcFile + "' is a directory");
10.   if (destFile.exists())
11. -    throw new IOException("Destination '" + destFile + "' already exists"); //在 1.4 版本引入
12. +    throw new FileExistsException("Destination '" + destFile + "' already exists"); //在 2.0 版本变更类型
13. }
```

（a） 版本变更  V1.4→ V2.0

```
1. public static void moveFile(File srcFile, File destFile) ...{
2. +    validateMoveParameters(srcFile, destFile);// //throw other three exceptions //在 2.7 版本移动其他异常的位置
3.     if (srcFile.isDirectory()) {
4.         throw new IOException("Source '" + srcFile + "' is a directory");
5.     if (destFile.exists())
6.         throw new FileExistsException("Destination '" + destFile + "' already exists");
7. }
```

（b）版本变更 **V2.0 → V2.7**

```
1. public static void moveFile(File srcFile, File destFile) ...{
2. -    validateMoveParameters(srcFile, destFile);// //throw other two exceptions
3. -    if (srcFile.isDirectory()) {
4. -        throw new IOException("Source '" + srcFile + "' is a directory");
5. -    if (destFile.exists())
6. -        throw new FileExistsException("Destination '" + destFile + "' already exists"); //在 2.9 版本整体重构
7. +    moveFile(srcFile, destFile, StandardCopyOption.COPY_ ATTRIBUTES);
8. }
9.
10. + public static void moveFile(File srcFile, File destFile, CopyOption... copyOptions) throws IOException {
11. +    validateMoveParameters(srcFile, destFile); //throw other three exceptions
12. +    requireFile(srcFile, "srcFile"); //throw other two exceptions
13. +    requireAbsent(destFile, null);
14. + }
15.
16. + private static File requireFile(File file, String name) {
17. +    Objects.requireNonNull(file, name);
18. +    if (!file.isFile())
19. +        throw new IllegalArgumentException("Parameter '" + name + "' is not a file: " + file);
20. +    return file;
21. + }
22.
23. + private static void requireAbsent(File file, String name) throws FileExistsException {
24. +    if (file.exists())
25. +        throw new FileExistsException(String.format("File element in parameter '%s' already exists: '%s'", name, file));
26. +}
```

（c）版本变更 **V2.7 → V2.9**

图 3  Apache Commons-IO 异常变更示例代码

## 3. Java 程序异常摘要提取方法

本章介绍异常摘要提取模块的主要方法。

### 3.1. 异常摘要提取模块概览

JavaExP 的异常摘要提取模块主要包含基本信息分析和异常前断言分析两个部分。如图 4 所示，**基本信息分析部分**以 jar 包或 class 文件为输入，负责构建程序的控制流图、函数调用图等数据结构，并提取每个方法中抛出异常的基本信息，获得方法到异常的映射。**异常前断言分析部分**首先通过构建方法的控制依赖图，去除判定结果与异常抛出行为无关的非控制依赖条件，提高了断言分析结果的精准性。获取依赖条件后，再通过参数约束推断将异常触发条件关联到外部输入参数，获取单个方法的前断言。此外，通过函数调用关系和参数传递关系追踪进一步构造了跨过程的异常抛出前断言。最后，形成异常信息摘要报告。

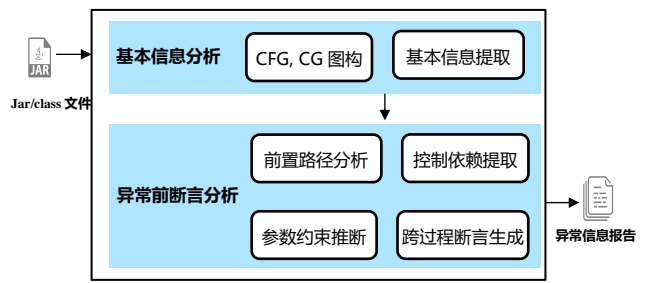

图 4  Java 程序异常摘要提取模块流程图

### 3.2. 基本信息分析

基本信息提取部分基于静态分析框架 Soot [44] 对输入代码进行预处理，为每个程序方法构建控制流图（Control Flow Graph，CFG），并生成全局的函数调用图（Call Graph，CG）。接着，通过遍历所有的语句，可以定位到显式抛出异常的 throw() 语句（称为异常抛出点），分析在每个异常抛出点抛出的异常类型和描述文本等异常基本信息。对于异常类型，我们分析异常变量的定义语句，并分析对应实例的类型；对于描述分析，我们从异常抛出

点的描述文本变量反向追踪，通过字符串函数建模还原完整的文本字符串，考虑到部分变量取值无法直接获取，这里将异常拼接后的文本信息转换为正则表达式形式。

获取基本信息后，JavaExP 记录方法名称和方法中显示抛出异常的映射关系，并生成方法-异常映射表，其中一个方法可以对应多个异常。对于图 3(c)的示例应用，可以得到一条映射边{ requireAbsent () → FileExistsException @loc25}。由于方法 moveFile 的参数会影响 requireAbsent 的异常抛出，因此对 moveFile 也应生成异常前断言。在后续分析中，该断言可通过为requireAbsent方法构建异常行为摘要和追踪跨过程参数关系得到。

### 3.3. 前断言分析

除了基本信息，异常摘要中的另一类重要元素是前断言信息。前断言分析包括控制依赖条件分析、外部输入参数约束推断和跨过程参数约束推断三个部分。

#### 3.3.1. 控制依赖条件分析

为了生成异常的前断言，首先需要准确提取异常的控制依赖条件。控制依赖条件一定在异常触发的前置路径上，即在异常被触发时经过的程序路径上。该路径可以从异常抛出点通过后向路径遍历得到，而该路径上的全部条件被称为为异常触发的前置路径条件。

**定义 5（异常前置路径）**：PrePath (m,e) = ($S_0$,…,$S_i$, $S_{i+1}$,..,$S_e$) 为异常触发的一条前置路径，其中 $S_0$ 为方法 m 的入口语句，$S_e$ 为异常 e 的抛出语句，程序语句 $S_{i+1}$ 是语句 $S_i$ 的一个后继节点。异常 e 对应的多个异常前置路径形成了异常前置路径集合 PrePathSet (m, e)。

**定义 6（异常前置路径条件）**： CondInPath (m, e, prePath) 为异常前置路径上的条件语句的集合，每个元素 cond ∈ CondInPath均是一个条件语句。

对于图 3(a)中的 moveFile () 方法，第 2-9 行分别抛出四个异常，如果他们的异常条件被满足，第 12 行异常 e 不会被抛出，因此，异常 e 依赖于这些控制条件。但如果将 2-9 行中的 throw 语句更改为非终止语句，如输出、日志、数据处理等语句，则其所属的条件将与第 12 行的异常无关。所有与异常抛出无关的路径前置条件语句无需被作为最终异常断言的一部分。

**定义 7（异常控制依赖条件）**：ControlCondInPath (m, e, prePath) 为异常前置路径上控制依赖条件的集合，其中每个元素 controlCond ∈ ControlCondInPath 是一条和异常 e 之间存在控制依赖关系的条件语句。即对于条件 controlCond 的多个后继节点，存在至少一个后继节点不存在于异常 e 的任何异常前置路径中。

**定义 8（异常控制依赖约束）**：ControlConstraintInPath (m, e, prePath)为异常前置路径上的异常控制依赖约束集合，每个异常控制依赖约束 (controlCond, isCondTrue) ∈ ControlConstraintInPath包括一个控制依赖条件语句及其条件判定结果。

算法 1（extractConstraint）给出了异常控制依赖条件的提取算法。第 2 行首先获取方法 m 的控制流图 cfg，其中节点代表语句，边代表语句之间的控制流向关系。第 3 行提取控制依赖图 cdg [32][33]，其中节点代表语句，边代表语句之间的控制依赖关系。通过搜索 cdg，可以得到异常抛出语句 $S_e$ 对应的异常控制依赖约束集合 controlConstraintSet。接着，第 5 行通过在 cfg 上后向路径遍历得到异常前置路径集合 prePathSet。对于 prePathSet 中的每条异常前置路径 prePath，第 7-13 行负责构建仅包含异常控制依赖条件切片的集合 controlConstraintInPath，并在第 14 行将其加入输出集合 controlConstraintSet 中。在这一过程中，第 8 行遍历 prePath 中的每一个节点，如果一个节点是条件语句，且存在于 $S_e$ 的控制依赖节点集合 controlNodeSet 中，将记录该节点 node（第 9 行）。第 10 行分析 node 在 prePath 上的后继节点 node.succ，判定当前路径上 if 条件的判定结果为 true 或为 false（在字节码中，if 语句会指明当 if 条件为真时的 goto 语句的位置，因此，可通过下一语句 succ 是否为 goto 的目标语句判定条件是否取值为真）。随后在第 11 行，该节点 node 与条件判定结果 isCondTrue 形成的约束条件 constraint 会被加入当前路径的异常控制依赖约束集合 controlConstraintInPath 中。最后，第 16 行将返回方法 m 的异常控制依赖约束集合 controlConstraintSet，其中每个元素为一条路径上的一组控制依赖约束。

**算法 1 控制依赖条件分析 extractConstraint**

**输入**：方法 m, 异常 e（异常抛出语句为 $S_e$）
**输出**：异常控制依赖约束集合 controlConstraintSet

1 Set <Set<Constraint>> controlConstraintSet = new

```
    HashSet (); //初始化
2   Graph cfg = constructCFG (m); //构建控制流图 cfg
3   Graph cdg = constructCDG (cfg, m); //构建控制依赖图
    cdg
4   Set <Node> controlNodeSet= getControlNodesOfExp
    (cdg, Sₑ); //在 cdg 中找到 Sₑ 的控制依赖节点
5   Set <Path> prePathSet = backTraverseFromExp (cfg,
    Sₑ); //从 Sₑ 后向路径遍历得到异常前置路径
6   for (Path prePath : prePathSet){
7       Set <Constraint> controlConstraintInPath= new
        HashSet ();
8       for (node : prePath) {
9           if (node.isCondition () and
            controlNodeSet.contains (node)) {
10              Boolean isCondTrue =getCondJudgeRes (node,
                node.succ); //获取分支条件的判定结果
11              controlConstraintInPath.add (new Constraint
                (node, isCondTrue)); //新增控制依赖条件
12          }
13      }
14      controlConstraintSet.add (controlConstraintInPath)
        //增加一条路径上的一组控制依赖条件
15  }
16  Return controlConstraintSet
```

例如，对于图 3(a)中的 moveFile() 方法，第 2-11 行的 5 个控制条件均存在一个后继节点不在任何一条异常前置路径中，因此异常控制依赖条件为 ControlCondInPath ={2,4,6,8,10}。接着，通过分析异常控制依赖条件在异常前置路径中的后继语句，即条件为真或为假时的后继语句在异常前置路径中，可以得到每个异常控制依赖条件取值结果，即 controlConstraintInPath = {(srcFile == null, false), (destFile == null, false), (srcFile.exists(), true), (srcFile.isDirectory(), false), (destFile.exists(), true)}。这里仅为方便展示，在实际的字节码分析过程中，可被获取的是中间变量约束，如($z0==0, false)。

### 3.3.2. 外部输入参数约束推断

经过上一节的分析，可以获取异常的控制依赖约束。但字节码中没有变量名称信息，仅有按序编号的内部变量，如 r0，z1，而这些变量在不同版本中不存在关联关系，难以被直接用于匹配和比较。因此，应将约束的主体转换为语义固定的对象，如 API 的参数，提取异常前断言时通过数据流追踪获取的内部变量约束归约为参数相关约束，从而准确分析异常前断言的变更情况。

算法 2（refineAnalysis）对于异常控制依赖条件集合中每条路径上的每个控制依赖约束条件 controlConstraint 进行分析，通过参数推断得到仅与外部输入参数相关的 refinedConstraint。算法第 1 行提取语句中的所有变量并将它们放入集合 dataRelatedVars。第 2-4 行中，对于变量集合中非输入参数相关的内部变量，通过数据流分析获取其最近的变量赋值语句，其中方法 getDefUse 通过数据流分析获取方法 m 的定义-使用链 [45]。第 5 行根据数据分析结果更改原约束条件 controlConstraint，即依次使用内部变量的赋值结果替换该变量，并将更新后的约束条件递归地传入 refineAnalysis 方法继续分析，直至 refinedConstraint 中不再包含内部变量时，迭代终止。这里，中间码预先被转换为 SSA 格式。此外，JavaExP 还通过启发式策略调整优化了输出形式，增强断言的用户可读性。该算法最后返回推断后的约束条件 refinedConstraint。

### 算法 2 参数约束推断分析 refineAnalysis

**输入**：方法 m, 异常 e（异常抛出语句为 Sₑ），异常控制依赖条件 controlConstraint

**输出**：推断后的约束条件 refinedConstraint

```
1  Set <Value> dataRelatedVars= getVarFromStmts
   (constraint.getStmt ()); //提取语句中的变量
2  For (Value: value dataRelatedVars){
3      if(isOutsideValue(value)) continue;
4      Stmt assignStmt= getAssignStmtofValue
       (getDefUse(m), value); //后向数据流分析获取最近
       的变量赋值语句
5      refinedConstraint = replaceValueInConstraint
       (constraint.getStmt (), value, assignStmt.
       getRightValue()) //将约束语句中的 value 替换为
       value 的赋值内容
6      refineAnalysis(m, e, refinedConstraint)
7  }
8  Return refinedConstraint
```

对于图 3(c) 中的 requireAbsent 方法，经过数

据流追踪，得到该方法内的控制依赖条件为 ($z0==0, false)}。表 1 (a) 中给出了通过数据流分析反向推断外部数据约束 parameter0.exists()的过程。

表 1(a) 过程内追踪约束条件 (requireAbsent)

| 控制依赖 | 数据流追踪后推断的约束条件 |
| --- | --- |
| ($z0==0, false) | $z0 is true + $z0 is-invoke r0.exists() |
| | → r0.exists() is true |
| | r0.exists() is true + r0 denote parameter0 |
| | → parameter0.exists() is true |

### 3.3.3. 跨过程参数约束推断

在 API 的演化过程中，异常的抛出位置可能发生移动，如将异常抛出语句移动到另一方法并调用它[17]。如采用过程间分析，这类代码重构会被识别为异常的删除，从而引发 API 演化分析中的误报。如图 3 对应的异常实例，在版本 1.4 中，异常抛出代码仅在 moveFile()本身中出现，但在最新的 2.13 版本中，该异常需要至少联合分析六个函数的才能被准确获取。为了增加异常匹配分析的准确度，JavaExP 在为过程内所有异常实例构造摘要信息的基础上，通过函数调用关系分析和参数映射关系分析，生成跨过程的异常摘要信息。

具体的跨过程参数约束推断的过程为：首先构造应用程序的函数调用图，并对函数调用关系进行拓扑排序。再按照拓扑序的逆序自底向上的分析每个方法，其中被调用的方法 callee 一定会比调用它的方法 caller 更早被分析。当定位到当前方法 caller 中的一个调用语句 stmt 时，可得到被调用的方法 callee。如果被调用方法 callee 的参数与异常的抛出有关，则会根据调用方法 caller 和被调用方法 callee 中参数位置的映射关系，更新从被调用方法 callee 中获取的参数相关约束 $cons_1$。此外，还应提取 caller 方法中调用 callee 语句前的程序路径上需要满足的参数相关约束，包含路径上的控制依赖约束 $cons_2$，和路径上被调用函数中的其他异常抛出约束的取反 $cons_3$。约束 $cons_1$、$cons_2$、$cons_3$ 均更新到调用方法 caller 的参数约束中后，可得到关于方法 caller 的异常前断言。对于函数内直接抛出的异常，类似的，也应使用异常抛出前置路径的函数调用中使得异常不被抛出的条约束 $cons_3$ 更新其直接约束。按照拓扑逆序分析，从而可以依次更新每个方法的前断言信息，由于底层的约束会向上传递更新，每个方法仅需被分析一次。

对于图 3(c)中的 moveFile 方法，其第 1 个参数对应方法 requireAbsent 中的第 0 个参数，根据这一参数映射关系可以更新 requireAbsent 中的前断言信息得到 moveFile 的约束。此外，在 moveFile 中调用 requireAbsent 方法时，执行到该方法调用点时应满足来自前置方法 validateMoveParameters 和 requireFile 中的其他约束，最终得到由 5 个约束组成的完整前置条件。结果见表 1(b)。

表 1 (b) 跨过程追踪约束条件 (moveFile)

| 控制依赖 | 数据流追踪后推断的约束条件 |
| --- | --- |
| constraint in *requireAbsent* + constraints in *moveFile* | requireAbsent@parameter0.exists() is true + requireAbsent(r1,null) |
| | → r1.exists() is true |
| | r1.exists() is true + r1 denote parameter1 |
| | → parameter1.exists() is true |
| | *merge constraints from related methods* |
| | parameter0 is not null + parameter1 is not null + parameter0.exists() is true + parameter0. isFile() is true + parameter1.exists() is true |

## 4. 基于异常提取的 API 生命周期构造

本章介绍 API 生命周期构造模块的主要方法。

### 4.1. API生命周期构造模块概览

JavaExP 的生命周期构造模块主要包含 API 匹配与变更分析、生命周期模型构造两个部分。如图 5 所示，**异常匹配与变更分析部分**以异常摘要文件为输入，先采用完全匹配策略获取 API 异常实例的匹配关系，再通过自适应匹配策略识别其他异常实例的映射和局部变更情况，生成异常敏感的 API 变更报告；**生命周期模型构造部分**则以多个版本的 API 变更分析报告为输入，分析同一 API 方法或同一异常实例在不同版本中的变更过程，最后生成相应的生命周期模型。

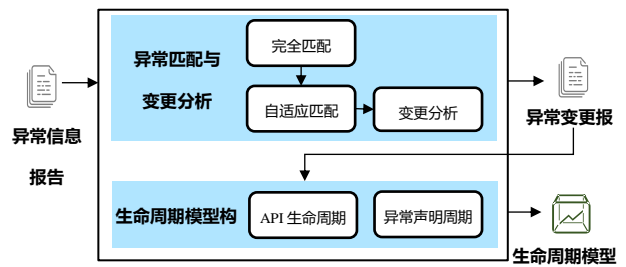

图 5 API 生命周期构造模块流程图

## 4.2. 异常信息敏感的API匹配与变更分析

基于提取的异常摘要信息,接下来,我们采用完全匹配和自适应匹配相结合的异常匹配方式对不同版本中的异常进行匹配。在异常匹配过程,API签名、异常类型、描述和断言信息是四个关键信息,这里仅考虑签名相同的 API 中异常的匹配,如果API 签名变更,则认为发生方法级改变。如表 2 所示,异常类型、描述和断言信息的不同组合共有 8 种。如果两个异常符合规则 R1,三个信息可完全匹配,则表明该异常未发生改变。如不能完全匹配,则需根据规则 R2-R8 进行自适应匹配。在这些规则中,我们将根据对符合该类别特征代码实例的经验分析和意图理解,判定不同变更特征下的异常代码是否可匹配。对于R2-R4,仅有单一信息变更,这类变化更有可能是同一异常的正常演化导致的,而非两个高度相似的异常。而如果多个信息变更,其匹配情况将相对复杂,如果两个异常摘要中有至少两类信息不一致(R5-R8),通常可认定两个异常指向不同的实例。但考虑到异常类型这一信息指向性较为稳定,在代码重构中有可能出现类型不变但是异常描述和断言发生改变的情况,因此规则 R5 下的匹配结果将视情况而定,具体参见以下异常匹配的详细流程。

**表 2 异常实例匹配规则**

| 规则 | API | 异常摘要 | | | 匹配结果 |
|---|---|---|---|---|---|
| | 签名 | 类型 | 描述 | 断言 | |
| R1 | √ | √ | √ | √ | 匹配-未变更 |
| R2 | √ | X | √ | √ | 匹配-类型变更 |
| R3 | √ | √ | X | √ | 匹配-描述变更 |
| R4 | √ | √ | √ | X | 匹配-断言变更 |
| R5 | √ | √ | X | X | 可能匹配-断言和描述变更 |
| R6 | √ | X | √ | x | 不匹配 |
| R7 | √ | X | X | √ | 不匹配 |
| R8 | √ | X | X | X | 不匹配 |

1) 对于所有待分析 API,采用规则 R1 对 API 在相邻版本中的异常进行完全匹配,如果匹配成功则将匹配到的一对异常分别从各自版本下的匹配队列中移除。
2) 采用异常类型变更 R2、异常描述变更 R3、和异常断言变更 R4 规则对应的三个规则进行匹配。这里将择依次使用各个信息过滤器,比较异常类型、描述、前断言信息是否一致。过滤器接受一组异常,根据过滤器类型返回过滤后的一组异常。如旧版本中的异常 $e_1$ 经过第一个过滤器之后匹配到多个异常,那么所有被匹配的异常会被作为候选对象传递给下一个过滤器;如 $e_1$ 未匹配到任何异常,则将当前过滤器接受到的全部异常将被作为候选对象传递给下一个过滤器。如果在过滤过程中,$e_1$ 在新版本中唯一匹配到异常 $e_2$,则表明 $e_1$ 和 $e_2$ 指向同一个异常,$e_1$ 和 $e_2$ 将被记录并从各自的匹配队列中移除。
3) 经过步骤 2 后未找到唯一匹配的异常,将继续采用规则 R5 进行匹配。满足这种情况的可匹配异常往往是由于代码重构造成的,由于前置方法、代码位置变更导致了异常前断言不完全相同,但这种情况下异常最近最直接的抛出条件通常不会改变,因此,可提取距离异常抛出点最近的直接条件对应的前断言(关键前断言)信息用于匹配。对于异常 $e_1$,如果能够唯一找到异常 $e_2$,它们的异常类型和异常关键前断言均相同,即使断言和描述信息不同,$e_1$ 和 $e_2$ 也会被匹配,记录后从各自的匹配队列中移除。
4) 经过步骤 3 后,待匹配异常集合规模可能减小,使得之前一对多匹配的异常变更为唯一匹配。因此,可再次重复过程 2)和 3),直至匹配结果不变,即达到不动点。
5) 如果多轮匹配结束后异常仍未被匹配,那么旧版本中的未匹配异常将被标记为异常删除,新版本中的未匹配异常将被标记为异常新增。

在图 3 代码中,JavaExP 根据规则 R2 匹配到V1.4-V2.0 的变化,根据规则 R1 完全匹配了 V2.0-V2.7,最后根据规则 R5 匹配到 V2.7-V2.9 的变化,从而完整刻画了该 API 中异常的变化情况。

## 4.3. 异常敏感的API生命周期模型构造

经过异常匹配与变更分析后,可得到 API 在相邻版本中异常的变更结果。变更类型包含**定义 4** 中声明的七类操作:API 新增、API 删除、异常新增、异常删除、异常类型变更、异常描述变更和异常断言变更。为了生成完整的生命周期报告,我们从初始版本开始,不断使用相邻版本的异常摘要构造更完整的生命周期报告,直至最新版本。

图 6 给出对图 3 示例中异常实例的生命周期报告。该报告包含异常所在的 API 的信息,如引入和删除版本;包括异常的基本信息,如异常的引入、删除版本;以及在不同版本区间上异常摘要信息的

变更情况，如异常类型变更。在这个实例中，异常所在的 API moveFile()，在 1.4 版本引入，在最新的 2.13 版本中依然存在。其中目标异常 e 共发生了三次变更，分别是：2.0 版本变更了异常类型；2.9 版本中更改了异常的描述文本；2.9 版本中更改了异常前断言中涉及到的方法调用。值得注意的是，在 2.7 版本中，虽然相关代码被开发者重构，但异常信息分析表明，其异常抛出行为未发生任何改变，因此，该版本未出现在变更报告中。

图 6  图 3 中 API moveFile()的生命周期模型

## 5. 工具实现与实验分析

基于本文提出的方法，我们实现了一种基于静态分析的 Java 程序 API 生命周期模型自动分析工具 JavaExP [46]。该工具包含约 10k 行 Java 代码和 1.4k 行 Python 代码，依赖于底层分析框架 Soot [44] 完成中间码提取、控制流图、函数调用图和控制依赖图[32]等数据结构的构建。为了评估本文方法的有效性和效率，本章在多个基准数据集上对工具 JavaExP 开展了一系列实验，主要研究问题如下：

- **RQ1**：在手工构造和真实 Java 项目上，JavaExP 能否准确高效地提取异常摘要信息？
- **RQ2**：在真实框架/库项目上，JavaExP 能否正确构造异常信息敏感的 API 的生命周期模型？
- **RQ3**：在真实框架/库项目上，异常信息敏感的 API 的生命周期模型有何特征？

### 5.1. 实验设置

为回答 RQ1，我们手工构造了一个包含常见异常抛出方式的基准测试集 ExceptionBench [51]，涉及到多种常见的 Java 特性。数据集中的六个类分别为：基本场景 Basic 类，其中包含无条件抛出、判空条件抛出、字符串取值条件抛出、字符串操作条件抛出、逻辑与/或条件抛出等；跨函数调用场景 MultipleCall 类，包括多种跨函数调用场景；多路径场景 MultiplePath 类，包含 if-else 分支路径和 for 循环路径；多个异常场景 MultipleThrow 类，包含同一函数内多个异常和跨过程调用导致的多个异常；类字段变量使用场景 FieldValue 类；和一个融合了多种场景的综合场景 Motivation 类；共包含 40 个异常。此外，本实验复用了 Nassif 等人构造的面向 Apache Commons IO [49]的异常断言标注集合[34]。为保证对比的公平性，本实验仅将目标项目的源文件和对应 Jar 包作为输入，而未将任何项目外的第三方库、JDK 等代码作为分析目标，排除未显式抛出的异常和代码实现不在项目中的异常后，该数据集共包含 392 个独立异常。此外，本实验还选取了六个广泛使用的 Java 项目来评估工具在这些真实项目上的分析性能。评估标准包括异常摘要数量和运行时间等。这里对每个项目的分析时间上限均被设置为 2 小时。为回答 RQ2 和 RQ3，我们以 RQ1 中六个真实项目为演化分析目标，根据这些项目在 Maven 仓库[53]中的代码发布情况，收集了共计 60 个历史版本 jar 包。随后，我们在这些版本上对真实项目中 API 的演化情况进行分析。对于同一项目，演化分析时仅考虑同级别变更版本。

在实验有效性评估的对比工具选择方面，由于尚没有对于异常信息敏感的 API 演化分析的直接对比工具，我们首先选择 Java 异常信息提取工具，对异常分析这一重要模块的精度进行评估，再人工检验 API 演化分析结果的有效性。经调研，工作 [21][22]可通过自然语言处理方式从 JavaDoc 文档或注释中提取异常信息，但由于文档/注释信息和代码信息常存在偏差，不能将文档信息中提取的异常信息作为演化分析的对象。工作[34] [54]通过机器翻译方法为异常抛出代码生成文档或通过自然语言处理技术自动生成测试用例，但其异常信息分析结果无法直接用于异常演化分析。基于此，本文选择了最新 SOTA 的 Java 异常前断言提取自动化工具 WIT [26]，评估该工具与 JavaExP 在异常提取方面的能力差异。WIT 工具通过静态分析解析目标

Java 项目源代码，构造控制流图并提取跨过程路径约束，结合约束求解技术获得异常相关的变量约束信息，最终提取出 Java 异常的类型、前断言、描述等信息，且其工具是公开可获取的。

### 5.2. 实验结果与分析

#### 5.2.1. 异常摘要报告有效性评估 (RQ1)

对于异常分析提取，分析工具提取到的异常都是真实存在的，其错误结果分为两种：未识别到真实的异常，对应为漏报 FN；识别到真实异常，提取了错误的异常摘要信息，而遗漏了正确的异常摘要信息，可被认为既属于误报，也属于漏报信息。

表 3 工具在 ExceptionBench 数据集上的有效性

| 工具 | TP | FP | FN | Precision | Recall | F1-Score |
|---|---|---|---|---|---|---|
| JavaExP | 39 | 1 | 0 | 0.98 | 0.98 | 0.98 |
| WIT | 30 | 5 | 10 | 0.86 | 0.75 | 0.80 |

表 3 分别给出本文 JavaExP 工具和异常提取工具 WIT 工具在 ExceptionBench 数据集上的有效性评估结果，包括 TP、FP、FN 数值，和根据公式 Precision=TP/(TP+FP)，Recall=TP/(TP+FN)，F1-Score=2×Precision×Recall/(Precision+Recall) 计算出的精确度、召回率和 F1 分数的结果。可以看到，JavaExP 成功为其中的 39 个异常生成了正确的异常摘要报告，其中 1 个误报是由于异常涉及复杂的数据值变更，导致前断言分析不准确。在 WIT 不能处理的 10（5+5）个异常中，5 个由于包含冗余且错误的约束或涉及复杂的数据值变更导致异常前断言提取结果有误，5 个涉及不支持的语法特性导致异常摘要未成功提取。在精确度、召回率和 F1 分数三个度量指标上，JavaExP 均优于 WIT 工具。

表 4 工具在 DScribe 数据集上的有效性分析

| 工具 | TP | FP | FN | Precision | Recall | F1-Score |
|---|---|---|---|---|---|---|
| JavaExP | 300 | 56 | 36 | 0.84 | 0.77 | 0.80 |
| WIT | 137 | 17 | 255 | 0.89 | 0.35 | 0.50 |

表 4 给出了工具 WIT 和 JavaExP 在公开基准测试集 DScribe[34]中异常上的有效性分析结果。在该数据集上，WIT 正确生成的异常摘要数量为 137。WIT 不能正确生成摘要的异常数量为 255（238+17），其中有 238 个异常的摘要为空，17 个异常的摘要信息不准确。与之相比，JavaExP 成功生成了 300 个正确的异常摘要报告。JavaExP 不能正确生成摘要的异常数量为 92（36+56），其中有 36 个异常的摘要为空，56 个异常的摘要信息不准确。虽然 WIT 的分析精确度略高于 JavaExP，但其召回率显著下降。与 WIT 相比，JavaExP 同时实现了较高的精确度和召回率，F1 分数与 WIT 比相对提升了 60%。

通过对 JavaExP 错误结果的分类分析，我们发现对于 56 个摘要信息不准确的异常，其中的 22 个异常受限于循环条件展开次数，11 个异常存在无法正确分析的复杂关系，10 个异常缺少部分正确路径，8 个异常中被调用函数 callee 中的前断言约束无法正确映射到调用函数 caller 的参数，3 个异常没有正确处理 try 语句中的异常抛出，2 个异常的前断言条件自相冲突。对于另外 36 个摘要结果为空的异常，导致精度损失的一个原因是复杂跨函数参数传递增加了分析难度。如图 3 所示，随着版本更新，框架/库开发者在重构的过程中倾向于将异常抛出代码进行包装以便复用，这间接增加了分析的复杂性。在 Apache Commons IO 项目 V2.13 版本中，调用链长度不少于 5 的共有 578 处，调用链长度不少于 10 的共有 50 处，而调用链中任意一处异常摘要信息误差均可能影响最终的匹配情况。此外，JavaExP 对字节码的静态分析能力也影响了分析结果，如在处理位运算代码的断言条件时尚存在偏差、对静态变量的取值使用初始赋值，循环条件仅展开 0 次和 1 次等，这些分析影响了前断言中部分条件的准确性。

表 5 工具在真实项目上的分析结果

| 项目名 | 版本 | LOC | WIT | | JavaExP | | |
|---|---|---|---|---|---|---|---|
| | | | 总摘要数 | 时间/秒 | 独立摘要数 | 总摘要数 | 时间/秒 |
| Commons IO | 2.6 | 9,984 | 297 | 3,903 | 268 | 1,285 | 44 |
| JGraphT | 0.9.2 | 15,660 | 142 | 119 | 176 | 4,17 | 52 |
| GraphStream | 1.3 | 48,535 | 142 | 431 | 342 | 2,087 | 160 |
| Guava | 19.0 | 70,250 | 2,347 | 6,133 | 221 | 3,891 | 112 |
| Nashorn | 1.8 | 83,728 | 177 | 1,759 | 949 | 3,446 | 355 |
| Android | 10.0 | 546,655 | 524* | 7,200 | 7,906 | 51,915 | 1743 |
| 合计/平均 | | 129,135 | 3,692 | 19,545 | 9,862 | 62,624 | 2,466 |

进一步，我们在六个真实项目上进行分析性能的评估。表 5 给出了所选项目的名称、版本和大小，其中大小为不含注释的 Java 源码行数。后五列对比了两个工具的结果中异常摘要数量和分析时间。其中，对于大型的 Android 框架代码，WIT 超时导致未完成分析，因此，仅统计其在两小时内生成的结果，并计算其中包含的异常摘要数量。对于六个被测项目，WIT 用时约 5.5 小时，提取的总异常摘要

数量为 3,692。与之相比,JavaExP 用时仅 0.7 小时,提取的独立摘要数量为 9,862,总摘要数量为 62,624,分析效率提高约 7 倍,提取数量显著增加。基于这一结果,我们进一步分析了异常数量的增加是由于 JavaExP 分析到了更多的独立异常,还是由于独立异常在跨函数调用过程中在不同调用路径中重复出现导致的。我们根据异常抛出方法、异常抛出语句位置信息进行去重后的异常数量计为独立异常的数量,独立摘要的数量少于总摘要数量。据统计,WIT 的总摘要数量(3,692)显著少于 JavaExP 的独立摘要数量(9,862),由此可知,与 WIT 相比,JavaExP 不仅提取出更多的异常摘要结果,且成功分析了更多的独立异常。

实验表明,JavaExP 提取的异常摘要数量显著多于 WIT,且用时更短,带来这一优势的主要原因有以下三个方面。首先,JavaExP 基于字节码分析,不受限于新的 Java 语法特性,分析范围更广,能够正确处理 Java 的各种复杂语法特性支持,其分析能力不受到代码形式的影响。其次,JavaExP 对分析规模的限制更少,该方法并没有对每条路径上的节点数、函数内联后的节点数量等做严格限制[26],而是在遍历控制流路径时先提取终止于异常抛出的语句,仅分析异常抛出行为相关的代码切片,分析范围更为聚焦。此外,JavaExP 采用自底向上构建函数摘要的方式,对每个函数不会被重复分析,带来了明显的效率优势。除了效率优势,JavaExP 的分析准确度也较高。JavaExP 通过提取异常的类型、描述文本、前断言三类核心信息对异常进行刻画,在异常前断言分析时主动忽略了与当前异常抛出无关的非控制依赖条件,但沿着函数调用链追踪异常抛出必要的关键前置条件,提高了异常分析的准确性,这也为后续在不同版本中匹配异常的演化信息打下了良好的基础。

> **RQ1 结论**:相比于现有的异常分析工具,JavaExP 能够更加准确地提取异常的摘要信息,在现有数据集上,将分析精度提高了约 60%;通过跨函数摘要合并策略,将分析效率提高了 7 倍,并显著增加了成功提取的异常摘要数量。

### 5.2.2. API 生命周期模型构造正确性 (RQ2)

在 RQ2 中,我们选取六个项目中发布历史版本数量最多(19 个)的 Apache Commons IO 项目,人工确认 JavaExP 在该项目上 API 演化分析结果的有效性。对于 API 和异常增删修改的七种形式,为了保证公平性,本实验选取对象具体的方法为:对于每个变更类别,首先根据变更实例数量对所在的 Java 类(class)文件进行排序。对于各个变更类型,根据其变更总数,从所有变更实例中按照均匀分布采样间隔地选择实例。考虑到异常前断言的数量相对较多,在选择时容易选择到因方法封装在不同上层方法被重复调用的异常实例,为增加多样性,避免确认相似的异常,该类别下会对异常调用链进行过滤,仅收集未包含相同异常抛出方法的实例。我们对每种类别均收集 10 个实例(不足 10 时按实际数量)。最终,对于 7 种变更类型,共收集了 63 个变更实例,结果见表 6,详细的人工确认报告见[46]。

对于过程内的分析,大量跨过程调用引入的异常均无法被分析,占异常总数的 94%;对于在当前方法中抛出的可被分析的异常,异常抛出之前的跨过程调用也会对前断言产生影响。在表 6 中,过程内分析评估时仅选择在当前方法中抛出的异常,如果只考虑当前方法内出现的约束条件,分析结果均正确,但如考虑其他方法调用带来的隐式约束,有 5 个断言信息变更行为正确,其中 3 个异常摘要前断言信息提取结果完全正确。

表 6  API 演化分析的正确性

| | 统计 | API 新增 | API 删除 | 异常新增 | 异常删除 | 异常文本修改 | 异常类型修改 | 异常条件修改 |
|---|---|---|---|---|---|---|---|---|
| 过程内 | 数量 | 10 | 10 | 10 | 10 | 10 | 3 | 10 |
| | 正确 | 10 | 10 | 10 | 10 | 10 | 3 | 10[5/3] |
| 跨过程 | 数量 | 10 | 10 | 10 | 10 | 10 | 10 | 10 |
| | 正确 | 10 | 10 | 10 | 10 | 9 | 10 | 10[10/5] |

与之相比,跨过程的分析则可以分析被调用函数中抛出的深层异常和前置函数中的隐式约束。受限于字节码静态分析,在异常条件修改变更结果中,5 个异常摘要前断言信息提取结果完全正确。5 个异常摘要提取结果不完全准确,但它们不影响对断言变更检测结果的正确性,如循环展开有限次和位运算约束结果不完全准确,但变更前后能够显著区分。在异常描述变更结果中,有一处错误的异常匹配。这是因为版本 2.9 中的代码被大幅重构,原异常的类型、消息、前断言均发生了改变,但函数中刚好存在另一个与原异常类型和关键前断言均相同的异常,从而导致它们被错误匹配。总体来看,跨过程策略下,演化分析的整体准确率达到 98%,跨过程分析能够捕获到其他函数内存在的异常及其断言条件,变更分析结果整体较为准确。后续 RQ3 中异常演化分析默认采用跨过程分析策略。

经人工总结，影响 API 中异常匹配的可能因素包括：1）在跨过程传递分析中，异常断言分析的精度和过程间参数约束的分析可能传递影响最终的匹配结果；2）开发者可能同时修改同一个异常的多个信息，导致难以通过单一变化严格限制匹配规则，需在尽量避免错误匹配的前提下，尽可能识别出存在差异的同一异常。3）目前仅能匹配文本相等和逻辑相等，如果开发者换用语义相同的不同 API，如!isFile()和 isDirectory()语义相同，因无法自动判断前后是否一致，会识别其为异常条件变更。如需判断语义一致性，需在后续研究中引入语义分析。

> **RQ2 结论**：JavaExP 能够基于提取的异常摘要，准确构建 API 的生命周期模型，其中跨过程分析策略更为准确。部分异常的前断言信息存在精度损失，但对 API 演化分析的影响不大。

### 5.2.3. API 生命周期变更结果分析 (RQ3)

对于 RQ2 中的六个项目，我们在 Maven 仓库中收集了各项目的共计 60 个历史发布版本，并在表 7 中统计了各个项目的 API 变更情况。所有版本的分析时间共计 30 分钟。对于 API 本身的变更，新增 API 数量较多，随着版本演化，API 的数量整体趋向于一直增加，但也有部分 API 会被删除。当发生大版本重构时，API 变化较为明显。除了 API 的新增删除，JavaExP 还识别出了大量的异常变更行为。在所有的 75,433 个 API 中，约 14.3%的 API 新增过异常抛出行为，13.9% 删除过原有的异常，6.5%更改过抛出条件，1.9%更改过异常描述文本，0.1%变更过异常类型。在异常敏感的 API 生命周期模型中，约 20%的 API 在被引入后，异常信息在后续版本中发生过调整，这说明 API 中异常相关代码的调整是十分常见的，异常敏感的 API 生命周期构造能够更加精准的描述 API 的实际变更情况。

进一步的，我们在图 7 和表 8 中展示了六个项目中异常实例的变更情况。图 7（a）中统计了每个 API 中的发生变更的全部异常实例，当一个异常被封装并多次调用时会多次统计。可以看到，当考虑重复异常时，新增异常的数量占比最高，其次删除异常的数量。实际上部分新增异常是被重复调用的。图 7（b）以独立异常为关注对象，异常多次调用时仅统计一次，表 8 给出了对应图 7（b）中独立异常变更的数量统计，与图 7（a）相比，新增删除异常的数量占比有所下降。在异常语义行为的不同变更中，异常类型变更整体数量最少，主要包括子类到父类的变更，父类到子类变更，原生异常类到自定义异常类的变更，代码重构复用功能相似代码导致的类型变更等。描述信息变更其次，其原因包括修正文字错误、增加描述信息、代码重构导致使用封装代码的描述文字等。而异常前断言更改的数量相对较多，包括增删路径上的前置异常导致条件变化，代码重构导致的条件变更等。

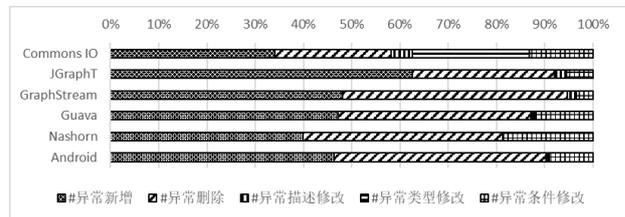

（a）异常实例变更类型统计

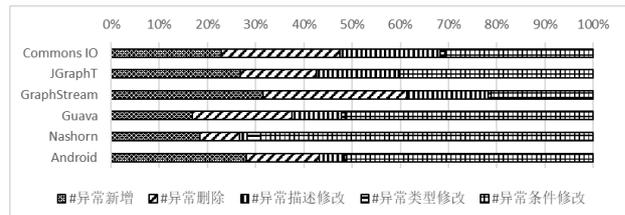

（b）独立异常实例变更类型统计

图 7 异常变更情况统计

表 7 API 变更情况统计分析

| 项目名 | #分析版本 | #总API | #新增API | #删除API | #异常摘要改变 API |
|---|---|---|---|---|---|
| Commons IO | 19 | 1,880 | 1,665 | 36 | 548 (29%) |
| JGraphT | 7 | 2,493 | 1,859 | 740 | 665 (27%) |
| GraphStream | 5 | 3,345 | 1,607 | 1,460 | 194 (6%) |
| Guava | 14 | 3,772 | 1,868 | 666 | 848 (22%) |
| Nashorn | 5 | 4,340 | 8 | 11 | 739 (17%) |
| Android | 10 | 59,603 | 32,331 | 8,260 | 11,512 (19%) |
| Total | / | 75,433 | 39,338 | 11,173 | 14,506 (20%) |

表 8 不同项目中独立异常变更统计

| 项目名 | #异常 | #异常新增 | #异常删除 | #异常类型修改 | #异常描述修改 | #异常条件修改 |
|---|---|---|---|---|---|---|
| Commons IO | 746 | 148 | 160 | 6 | 135 | 200 |
| JGraphT | 979 | 87 | 51 | 1 | 55 | 130 |
| GraphStream | 485 | 58 | 55 | 1 | 31 | 39 |
| Guava | 453 | 46 | 56 | 2 | 28 | 140 |
| Nashorn | 1,332 | 72 | 32 | 11 | 6 | 269 |
| Android | 11,471 | 1,669 | 901 | 21 | 310 | 3,039 |

**RQ3 结论**：在代码演化过程中，不仅 API 的新增、删除行为较为常见，异常的新增、删除和更改行为也十分频繁。在 75,433 个被分析的 API 中，约有 20% API 的异常抛出行为至少发生过一次改变，这些 API 共涉及超过七千多处独立的异常变更。相比于 API 的存在性生命周期模型，采用异常敏感的分析时，API 发生变动的比例提高了 18%，该模型能够更加精准地描述 API 的实际变更情况，对框架/库代码的开发者和使用者都具有指导意义。

#### 5.2.4. 对实验有效性的威胁

在本文中，对实验有效性的威胁主要与数据集的构建与选取有关。1）在手工基准测试集的构建阶段，设计思路的不同会在一定程度上影响在该数据集上的评估结果，这一偏差难以避免。为了保障测试集的公平性，本文在构建手工基准测试集时预先对抛出异常的基本场景进行分类，再按照类别设计数据集代码。对于这些场景，该数据集仅考虑具有指定特性的精简代码片段，用于测试异常分析工具在具有不同特性代码上的基础分析能力。我们注意到 WIT 不支持一些常见的语法特性，为保障公平性，我们仅根据数据集设计需要设计不同特性的代码片段，而未在设计后主动添加移除特定工具（如 WIT）不支持的语法特性。2）在真实项目数据集的选取上，真实项目的异常抛出代码风格、函数封装复杂度、相邻版本代码变更差异等均会影响评估结果，为增强被测项目的代表性，本文复用已有的异常断言标注集合[34]用于异常分析能力评估，并选取了六个广泛使用且具有多个版本的 Java 开源项目用于 API 变更分析评估，通过工具在不同被测项目上的整体结果评估工具的综合分析能力。

## 6. 相关工作

### 6.1. 异常摘要提取

异常（Exception）机制是 Java 中正式的错误报告机制，为了能够及时有效地处理程序中的运行错误，开发者需合理地抛出、捕获并处理异常。程序崩溃时，打印的异常堆栈信息是错误调试的一类重要信息[57]。由于异常机制的复杂性，研究人员围绕着异常的使用[38]、异常抛出代码的编程指导[39][40]、异常抛出行为的正确性测试[37][56]、以及基于程序异常抛出信息的错误定位与修复[47][46][48]等方向开展了一系列工作。其中，为了帮助开发人员了解代码中何时何处会抛出异常，理解程序的规范行为，异常的摘要信息，特别是其前断言生成工作也受到了广泛的关注，主要类别包括基于自然语言处理的方法和基于代码静态分析的方法。

#### 6.1.1 基于自然语言处理的断言提取

在开发过程中，断言信息可以帮助开发人员明确方法的使用规范，避免 API 演化导致的代码缺陷；在代码缺陷检测和定位时，前断言分析结果可以辅助测试人员构造高质量的测试用例，对满足/不满足断言的行为进行系统地测试。

基于自然语言处理（NLP）的断言生成方法被广泛地应用，这类方法通过统计分析文档、注释等文本文件推断方法的断言或测试预言信息[5][20][21][22][23][24]。Tan 等人在@Tcomment[1]中通过定义自然语言模式和使用启发式方法来推断程序的异常前断言，该方法仅关注空指针类型。与之相比，Goff 等人提出的 ToraDocu[21] 通过解析 Javadoc 文档，自动为所有的异常行为构造测试预言，工作 JDoctor[22] 在此基础上扩展，实现了面向更多程序行为的断言提取。此外，Zhai 等人提出了从文档中自动生成 JML 规范的方法 C2S[20]。由于大部分真实应用并不存在完整的 JML 规范，该方法仅基于 JDK 的规范文档进行训练，其模型不一定适用于其它的 Java 项目。基于自然语言处理的方法可以有效基于文本分析实现断言提取，但无论是方法文档或是代码注释，开发者对它们的编写情况都是不确定的。代码中的文档、注释信息既可能缺失，也可能在代码演化过程中未被及时更新，因此，这类方法适用于文档编写较为规范且被长期维护的大型项目。但在大量真实项目中，存在着文档缺失、不完整或未被及时维护的现象[25][35][36]，无法准确体现 API 代码实现本身的演化情况。

#### 6.1.2 基于代码分析的断言提取

另一类方法基于静态代码分析来提取断言信息[52]。Buse 和 Weimer 基于 Java 异常分析工具 Jex[30] 提出了一种自动推断 Java 方法异常抛出条件的方法 [29]。该工作首先提取方法和异常的映射表，然后采用符号执行和跨过程数据流分析技术提取每个异常的抛出条件。由于该工作后向遍历了所有的控制流路径（control flow path），在单个方法代码复杂、异常抛出前存在分支条件较多的情况下会出现路径爆炸问题；收集到路径约束后，该方法

设计了一些约束处理规则以简化断言形式，但其生成的结果均是围绕所有程序变量的，而不是只关注异常和方法输入参数的关系。与之相似，Chandra 等人也采用后向符号执行技术，提出了一种推断最弱前断言的技术 SnuggleBug [31]，它将问题泛化为如何找到从某入口点到达目标状态的前置条件，因此该方法不限于异常分析。为了解决 Java 多态虚函数调用关系分析带来的路径爆炸问题，该工作采用符号执行和函数调用图交错的按需分析方法以提高效率。但它们[29] [31]均未提供可公开获取的工具。

近期，Marcilio 等人[26] 提出了基于 Java 源码分析的轻量级异常前断言分析方法 WIT。该方法可以有效提取部分 Java 方法断言，但由于面向源代码，其分析而受限于复杂的语法特性，如不能处理包含 for-each 循环语句, switch 语句, 和 try/catch 块的代码；此外，基于源码的分析依赖于变量名称的匹配，在变量重新赋值时难以准确解析条件变量和输入参数的关系。WIT 项目的源码未公开，但工具可公开获取。

在基于代码分析的断言提取方法中，于源码的分析可以有效提取部分方法的断言，但其受限于复杂且不断更新的 Java 语法特性、难以准确追踪内部变量和外部参数之间的复杂关系[26]。此外，对于上层应用依赖的底层框架和第三方库，其源代码未必是可获取的。与之相比，基于字节码的分析不会受限于高级语言的语法特性，并支持开展精确的控制流和数据流追踪。为增加方法的普适性，适应不同版本的 Java 代码，并支撑框架/库源码不可获取的分析场景，JavaExP 向 Java 字节码的静态分析技术，通过追踪分析字节码中的异常抛出条件和变量取值，实现异常相关的 API 语义变更分析。

### 6.2. API生命周期模型构建

API 生命周期模型常被用于上层应用的 API 误用检测或兼容性错误检测。Li 等人在工作 CiD [5] 中提出了安卓生命周期模型 (ALM)，CiD 从安卓开发框架中提取了完整的 API 方法列表，并给出不同 API 存在的版本范围。Huang 等人提出了 CIDER[8],该工作关注 API 回调函数变化导致的兼容性问题，该工作依赖于手工构建的回调函数调用协议一致性图。工作 ACID[7]同时关注 API 调用问题和 API 回调函数兼容性问题，该工作没有分析框架代码，而是根据安卓框架官方提供的 API 差异列表轻量级地获取其生命周期。本文也关注与框架 API 生命周期的提取，与这些工作相比，我们不仅关注 API 的存在性问题，即在不同版本中 API 的新增和删除情况，还重点分析了 API 中异常抛出情况，特别是同一个 API 中异常抛出条件、描述、类型等是否发生变化。

除了 API 方法的演化，工作[11]还关注了框架代码中字段（Field）信息的演化，并关注了字段变化引发的上层代码缺陷。更多的工作[6][9][10]关注与在给定 API 生命周期模型的基础上，如何精确分析上层应用代码，以找到 API 的误用问题，我们的模型提取工作可以为这类研究提供支撑。

## 7. 总结与展望

针对框架/库项目和上层应用开发者在代码升级演化过程中难以准确获取其开发或使用的 API 变更行为这一问题，本文基于静态分析方法，提出了面向底层框架和第三方库的异常信息敏感的 API 生命周期模型生成方法，形成原型工具 JavaExP。与已有工作相比，JavaExP 生成的异常摘要信息在准确率和分析效率方面均有大幅提高。与异常不敏感的 API 演化分析相比，异常敏感的 API 发生变动的比例提高了 18%，在六个真实框架/库项目的 60 个版本中发现了超过七千多处独立的异常变更。

这一工具可同时服务于框架/库的开发人员和使用人员。一方面，对于框架/库的开发者，应在发布新版本软件前，通过 API 生命周期分析工具精确获取新版本代码中 API 中异常信息变更情况，确保小版本升级时不产生 API 语义变更，大版本升级时及时将语义变更情况更新在文档中。另一方面，对于应用开发者，在对所使用的框架/库代码进行版本升级时，可通过分析工具查看当前版本到新版本中 API 的方法变更和异常变更情况，开展未捕获的异常分析和异常传播分析等应用层检测，保障应用层 API 调用的正确性和鲁棒性，服务于软件供应链安全分析。此外，对于基于大模型的自动代码生成，框架/库 API 误用是生成代码中的一种典型错误模式，API 生命周期信息对于生成代码的版本一致性检测和修复也有重要意义。

考虑到真实的大规模框架/库中异常信息变动非常频繁，在后续的研究工作中，我们将进一步探索如何从大量的异常信息变更中自动识别出可能影响代码可靠性的语义变化、如何自动构造可触发 API 中的异常抛出行为的测试用例等研究问题，从而精准定位上层软件系统中的 API 误用行为。

**参考文献：**